\documentclass{PoS}
\usepackage{psfig,epsfig,colordvi,amssymb,amsmath}

\def\lsim{\mathrel{\rlap{\lower4pt\hbox{\hskip1pt$\sim$}}
    \raise1pt\hbox{$<$}}}                
\def\gsim{\mathrel{\rlap{\lower4pt\hbox{\hskip1pt$\sim$}}
    \raise1pt\hbox{$>$}}}                

\newcommand{\be}{\begin{equation}}
\newcommand{\ee}{\end{equation}}
\newcommand{\bea}{\begin{eqnarray}} 
\newcommand{\eea}{\end{eqnarray}}

 \newcommand{\csw}{\, c_{\rm SW}}  
   
\newcommand{\Dlr}{\mbox{\parbox[b]{0cm}{$D$}\raisebox{1.7ex}
                       {${\,\scriptstyle{\leftrightarrow}}$}}}

\title{Perturbative renormalization of GPDs to $O(a^2)$, for various fermion/gluon actions}

\ShortTitle{Perturbative renormalization of GPDs for various actions}

\author{\speaker{Martha Constantinou}
\thanks{Work supported in part by the Research Promotion Foundation 
of Cyprus (Proposal Nr: TEXN/0308/17)}\,\,, 
Haralambos Panagopoulos, Fotos Stylianou$^\dagger$ \\
        Department of Physics, University of Cyprus,\\
P.O.Box 20537, Nicosia CY-1678, Cyprus\\
        E-mail: \email{marthac@ucy.ac.cy},\,\, \email{haris@ucy.ac.cy},\,\, \email{fstyli01@ucy.ac.cy} }

\abstract{We present a 1-loop perturbative calculation of the fermion
propagator, up to ${\cal O}(a^2)$ ($a$: lattice spacing). The fermions are
described by Wilson, clover and twisted-mass actions; for gluons we
use Symanzik improved actions (Plaquette, Tree-level Symanzik,
Iwasaki, TILW, DBW2). Our results are given in a general covariant
gauge, and their dependence on the coupling constant, the external
momentum, the masses and the clover parameter is shown explicitly. We
also study the ${\cal O}(a^2)$ corrections to matrix elements of
unpolarized/polarized fermion bilinear operators, which include up to
one derivative. These corrections are essential ingredients for
improving, to ${\cal O}(a^2)$, the renormalization constants of the
operators under study. In addition, they can be used to minimize
lattice artifacts in non-perturbative studies. 

\bigskip

}

\FullConference{The XXVII International Symposium on Lattice Field Theory\\
		 July 26-31, 2009\\
		 Beijing, China}

\begin{document}

\section{Introduction}

A major issue of Lattice Gauge Theory has been the reduction of
effects which are due to the finite lattice spacing $a$, in order to
better approach the elusive continuum limit. Over the years, many
efforts have been made for ${\cal O}(a^1)$  improvement in lattice
observables, which in many cases is automatic by virtue of symmetries
of the fermion action. According to Symanzik's program
\cite{Symanzik}, one can improve the action by a judicious addition of irrelevant
operators. Also, in the twisted mass formulation of QCD~\cite{FGSW} at
maximal twist, certain observables are ${\cal O}(a^1)$ improved, by
symmetry considerations. 
The first 1-loop perturbative computation of ${\cal O}(a^2)$ effects was
recently performed by our group~\cite{CPS}. This regards the
evaluation of the fermion propagator and Green's functions for
ultralocal bilinears of the form ${\cal O}^\Gamma_\alpha=\bar\Psi\Gamma
\lambda_\alpha\Psi$, using the Wilson/clover fermions and
Symanzik improved gluons. Extending a calculation up to ${\cal
  O}(a^2)$ brings in new difficulties, compared to lower order in $a$;
for instance, there appear new types of singularities. The procedure to
address this issue is extensively described in Ref.~\cite{CPS}. 
The ${\cal O}(a^2)$ terms of such perturbative computations are of
great utility since they can be subtracted from non-perturbative
estimates to minimize their lattice artifacts.

The generalized parton distributions (GPDs) of the nucleon determine
non-forward matrix elements of gauge invariant light cone operators;
the moments of such operators can be evaluated on the lattice.
GPDs also give information on interesting quantities such as the quark
orbital angular momentum contribution to proton spin~\cite{Ji}.
Moreover, the generalized parton distributions can be measured in high
energy scattering experiments, for instance the deeply virtual Compton
scattering of virtual photons off a nucleon. Thus, by computing moments of
GPDs in lattice QCD one can explore many aspects of the nucleon structure. 
The operators that are related to the GPDs must be renormalized,
before one compares results from simulations to physical,
experimentally measurable quantities. 

In this work, we investigate the perturbative renormalization of 
the fermion propagator, local and twist-2 fermion operators. We
compute all matrix elements for the amputated Green's functions of the
inverse fermion propagator, ultralocal bilinears and twist-2
one-derivative operators, defined as $\bar\Psi\Gamma_{\{\mu} 
\overleftrightarrow D_{\nu\}}\tau^\alpha\Psi$ (symmetrized and
traceless). Although our expressions for all these matrix elements are
as general as possible (the dependence on almost all parameters is shown
explicitly), these are extremely lengthy and complicated to be
presented here. Thus, we show our results only for the renormalization
constants and for particular choices of the various parameters. 
In Section~\ref{method} we briefly describe the procedure for these
computations and in Sections~\ref{prop} - \ref{twist2} we present our
results for the renormalization functions. We also compare with
non-perturbative estimates provided by the ETM Collaboration. 
Due to space limitations the renormalization of the ultralocal
operators is not discussed here, but will appear in a longer write-up.

\section{Description of the calculation}
\label{method}

In the framework of this computation we employ a fermion action which
includes the clover term, and also an additional mass parameter, $\mu$
\bea
\hspace{-0.5cm}
S_F &=& \sum_{f} \sum_x \bar{\Psi}_{f}(x)\Bigg[\frac{\gamma_\nu}{2}\left(\nabla_\nu+\nabla_\nu^\ast \right) -
\frac{a\,r}{2}\nabla_\nu^\ast\,\nabla_\nu + m + i\,\mu\gamma_5\tau^3 
- {1\over 4}\,c_{\rm SW}\,\sigma_{\rho\nu} {\hat F}_{\rho\nu}(x) \Bigg]\Psi_{f}(x) 
\label{action}
\eea
A summation over the Dirac indices $\nu,\,\rho$ is implied.
The advantage of the above action is that it combines different
actions that are used widely: Wilson ($\csw\to0,\,\mu\to0$),
clover ($\mu\to0$) and twisted mass ($\csw\to0$). It is
important to mention that for the case of twisted mass fermions,
Eq.~(\ref{action}) corresponds to the twisted mass action in the so
called twisted basis $\chi$ ($\chi=e^{-i\frac{\omega}{2}\gamma_5\,\tau^3}\,\Psi$).
The appropriate form of this action in physical basis has a rotation
of the Wilson term, $r \to - i\,r\,\gamma_5\,\tau^3$. This is crucial
because the definition of the $Z$-factors in physical and twisted
bases are different, for instance $Z_A^{phys} = Z_V^{twisted}$ and
$Z_V^{phys} = Z_A^{twisted}$, for the axial and vector currents with
or without a derivative. The results presented here are all in
the physical basis.

For the gluon part we use a family of Symanzik improved gluons and
more precisely 10 different sets of Symanzik parameters that are
widely used in simulations; these are shown in Table 1 of
Ref.~\cite{CPS}. In the same reference, the reader can find all
details for the calculation procedure and the evaluation of the
superficially divergent integrals.
The substantial difference with our previous computations is the
presence of masses, $m$ and $\mu$, in the fermion propagators. The
dependence of superficially divergent integrals on $p,\,m,\,\mu$ is
nontrivial and it is a rather complicated task to extract this
dependence explicitly. In fact, there are 11 "strongly" IR divergent
integrals (they are convergent only at $D>6$ dimensions) and a few
hundred of divergent integrals at $D=4-2\epsilon$ dimensions.

The 1-loop diagrams that enter our three calculations are shown in
Fig. 1. Diagrams 1a and 1b regard the correction to the inverse fermion
propagator, diagram 2 corrects the local bilinears and finally
diagrams 3a-3d contribute to the calculation of the twist-2 bilinears.
The appropriate operator employed in each computation is represented by a
cross in diagrams 2, 3a-3d.

\begin{center}
\psfig{figure=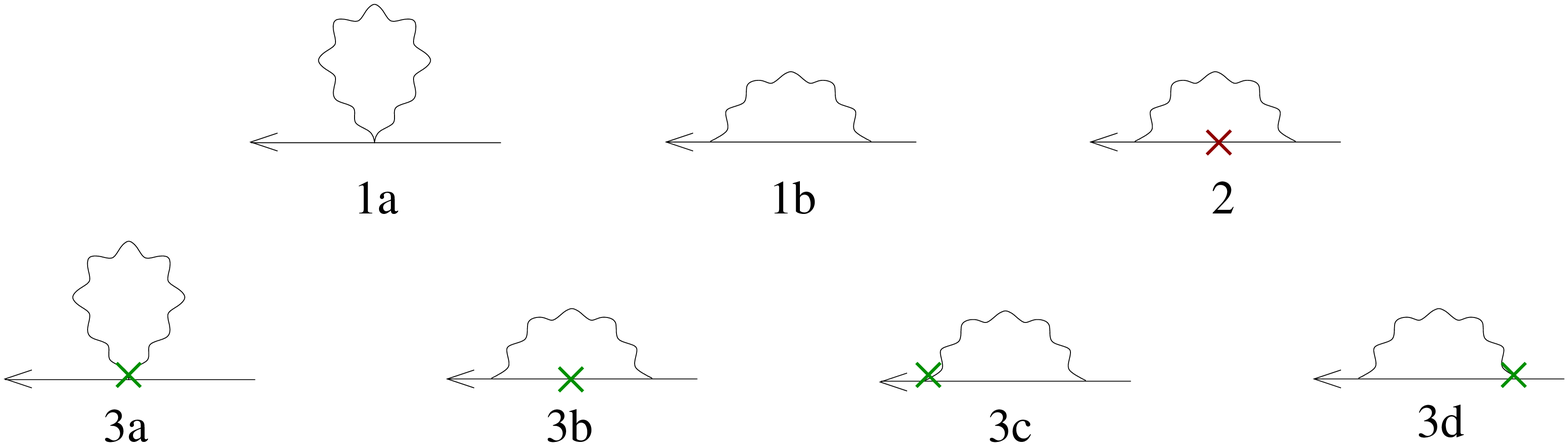,height=4truecm}
\end{center}
{\small 
\begin{center}
\begin{minipage}{14cm}
{\bf Figure 1:} One-loop diagrams contributing to the correction of the
amputated Green's functions of the propagator (1a, 1b), local bilinears
(2) and twist-2 operators (3a-3d). A wavy (solid) line represents gluons
(fermions). A cross denotes an insertion of the operator under study.
\end{minipage}
\end{center}}

\section{Renormalization of the fermion field to ${\cal O}(a^2)$}
\label{prop}

For all the renormalization factors used here we employ the RI'-MOM
scheme. The fermion field renormalization constant, $Z_q$, can be
obtained using alternative renormalization conditions that differ in
their lattice artifacts. The most widely used conditions are the
following two
\bea
\label{ZqA}
Z_q^A &=& \frac{1}{4} {\rm Tr}\Big[S_{\rm cont}(p) \cdot S^{-1}_{\rm
    1-loop}(p) \Big] \Bigg{|}_{p^2 = \bar\mu^2}\,, 
\qquad S_{\rm cont}(p)= \frac{-i\,\sum_\rho p_\rho\,\gamma_\rho}{p^2}\\
\label{ZqB}
Z_q^B &=& \frac{1}{4} {\rm Tr}\Big[S_{\rm tree}(p) \cdot S^{-1}_{\rm
    1-loop}(p) \Big] \Bigg{|}_{p^2 = \bar\mu^2} \,, 
\qquad S_{\rm tree}(p)= \frac{-i\,\sum_\rho\,\frac{1}{a}\sin(a\,p_\rho)\,
\gamma_\rho}{\frac{1}{a^2}\sum_\rho\sin^2(a\,p_\rho)}
\eea
with the trace taken over the Dirac indices.
The $S^{-1}_{\rm 1-loop}(p)$ is the amputated Green's function for the
inverse propagator up to 1-loop; it was computed from diagrams 1a and 1b of Fig. 1.  
Indeed, the constants $Z_q^A$ and $Z_q^B$ differ only in their lattice
artifacts, since the ${\cal O}(a^2)$ expansion of $S_{\rm tree}(p)$ is
simply $S_{\rm cont}(p) + i\,a^2 \,\sum_\rho p^3_\rho\,\gamma_\rho/(6\,p^2)$.

In the most general expression for $Z_q$, we show explicitly the
dependence on the action parameter $\csw$, the coupling constant $g$,
the number of colors $N_c$, the gauge fixing parameter $\lambda$, the
masses $m,\,\mu$, the lattice spacing $a$ and the external momentum
$p$. On the contrary, we cannot express $Z_q$ in a closed form as a
function of the Symanzik parameters, $c_0,\,c_1,\,c_2,\,c_3$; we
have computed $Z_q$ for each of the 10 sets of Symanzik coefficients
shown in Table 1 of Ref.~\cite{CPS}. Next we provide our result for
$Z_q$ using the condition in Eq.~(\ref{ZqA}), for the Landau gauge,
$\csw=0$, $m=0$ and tree-level Symanzik improved gluons
{\small{
\bea
Z_q^A = \Red{\Bigg[}&1& - \frac{\Orange{a^2}\,p4}{6\,p^2} +  \frac{C_F\,g^2}{16\,\pi^2}\Green{\Bigg\{}-13.0232726(2) 
+ \Orange{a^2}\,\Blue{\Bigg(}
-\ln[a^2\mu^2 +a^2p^2]\Big(\,\frac{2}{3}\,\mu^2
+ \frac{73}{360}\,p^2 
+ \frac{157}{180}\,\frac{p4}{p^2} \Big)  \nonumber\\ [1.5ex]
&+& 1.1590439(1)\,\mu^2 + 4.2770447(3)\,\frac{p4}{p^2} + 1.1471634(1)\,p^2 
+ \frac{7}{40}\,\frac{\mu^8}{(p^2)^3} \nonumber\\ [1.5ex]
&-& \frac{1}{240}\,\frac{\mu^6}{(p^2)^2} 
- \frac{37}{180}\,\frac{\mu^4}{p^2} 
- \frac{7}{20}\,\frac{\mu^8\,p4}{(p^2)^5} 
+ \frac{1}{120}\,\frac{\mu^6\,p4}{(p^2)^4} 
+ \frac{169}{180}\,\frac{\mu^4\,p4}{(p^2)^3} 
- \frac{43}{80}\,\frac{\mu^2\,p4}{(p^2)^2} \nonumber\\[1.5ex]
&+& \ln[1 + \,\frac{p^2}{\mu^2}]\Big(\,
- \frac{7}{40}\,\frac{\mu^{10}}{(p^2)^4} 
- \frac{1}{12}\,\frac{\mu^8}{(p^2)^3} 
+ \frac{2}{9}\,\frac{\mu^6}{(p^2)^2} 
- \frac{1}{3}\,\frac{\mu^4}{p^2} \nonumber \\[1.5ex]
&&\phantom{+\ln[1 + \,\frac{p^2}{\mu^2}]} 
+ \frac{7}{20}\,\frac{\mu^{10}\,p4}{(p^2)^6} 
+ \frac{1}{6}\,\frac{\mu^8\,p4}{(p^2)^5} 
- \frac{35}{36}\,\frac{\mu^6\,p4}{(p^2)^4} 
+ \frac{1}{12}\,\frac{\mu^4\,p4}{(p^2)^3} \Big) 
\,\Blue{\Bigg)}\,\Green{\Bigg\}}\,\Red{\Bigg]}\phantom{\Bigg{|}}_{\phantom{\Big{|}}_{\hspace{-0.2cm}p^2 =  \bar\mu^2} }
\label{ZqLandaySymanzik}
\eea}}
In the above expression $p4\equiv\sum_\nu p_\nu^4$,
$C_F=(N_c^2-1)/(2\,N_c)$, and $\bar\mu$ is the renormalization
scale. No ${\cal O}(a^1)$ terms appear in Eq.~(\ref{ZqLandaySymanzik})
since we set $m$ equal to zero.
Although the mass dependence shown in Eq.~(\ref{ZqLandaySymanzik}) is
very complicated, the numerical result for the renormalization
constant $Z_q$ depends mildly on $\mu$ (for the values used in
numerical simulations by ETMC, $0.003\leq\mu\leq0.01$). 

A very important issue is that the ${\cal O}(a^2)$ terms depend not
only on $p^2$, but also on the direction of the momentum $p$, as
manifested by the presence of $p4$. As a consequence, different
renormalization prescriptions, involving the same renormalization
scale $\bar\mu$ but different directions of $p$, lead to a different
renormalization constant. The strategy one may follow is to average
over the results for $Z_q$ which are obtained using these different
directions.

In Table 1 we present our 1-loop results for $Z_q$, up to ${\cal O}(a^2)$,
for the two renormalization conditions: Eq.~(\ref{ZqA}) ($Z_q^A$) and
Eq.~(\ref{ZqB}) ($Z_q^B$). The results are given in the chiral limit
for the Landau gauge, $\csw=0,\,N_c=3$ and for 3 values of the coupling
constant $\beta=2N_c/g^2$. Instead of the bare coupling we have used
the tadpole-improved coupling~\cite{LM}, defined as $g_t^2=g^2/\langle plaq\rangle$, 
to achieve further improvement. The renormalization scale $\bar\mu$
was set to $1/a$, that is $a^2\,p^2=1$. This is a well-defined choice,
since all possible lattice momenta (permutations of $(\pm1,0,0,0)$) have
the same $p4$. Thus, $Z_q$ is the same for all these permutations.
$Z_q^{A\,{\cal O}(a^2)}$ and $Z_q^{B\,{\cal O}(a^2)}$ are the ${\cal
  O}(a^2)$ effects of $Z_q^A$ and $Z_q^B$ respectively.
The last column gives the non-perturbative
estimates of ETMC~\cite{ETMC}, using the renormalization
condition of Eq.~(\ref{ZqB}) and employing the same parameters as
we did. In $Z_q^{non-pert}$ our ${\cal O}(a^2)$ corrections were
already subtracted to reduce lattice artifacts. In fact, ETMC applied
the subtraction procedure to all their data; therefore,
the behavior of $Z_q^{non-pert}$ against the renormalization scale becomes flatter in the
energy region where perturbation theory is valid. This indicates that the
lattice artifacts are suppressed.
\vspace{0.1cm}
\begin{center}
\begin{minipage}{15cm}
\begin{tabular}{lr@{}lr@{}lr@{}lr@{}lr@{}l}
\hline
\hline
\multicolumn{1}{c}{$\beta^{\phantom{A^{\phantom{A}}}}_{\phantom{A_{\phantom{A}}}}$}&
\multicolumn{2}{c}{$Z_q^A$} &
\multicolumn{2}{c}{$Z_q^{A\,{\cal O}(a^2)}$} &
\multicolumn{2}{c}{$Z_q^B$} &
\multicolumn{2}{c}{$Z_q^{B\,{\cal O}(a^2)}$}&
\multicolumn{2}{c}{\hspace{-0.2cm}$Z_q^{non-pert}$}  \\
\hline
\hline
3.80  &0&.655255014(4)  &-0&.039554503(1)  &0&.771056601(4) &0&.076247083(1) &\hspace{-0.2cm}0&.755(4)$^{\phantom{A^{\phantom{A}}}}$\\
3.90  &0&.663871997(4)  &-0&.045705298(1)  &0&.782134880(4) &0&.072557584(1) &0&.757(3)                                      \\
4.05  &0&.675276827(4)  &-0&.053846057(1)  &0&.796797308(3) &0&.067674424(1) &0&.777(5)$_{\phantom{A_{\phantom{A}}}}$                \\
\hline
\end{tabular}
\end{minipage}
\end{center}
{\small 
\begin{center}
\begin{minipage}{14.5cm}
{\bf Table 1:} Perturbative and non-perturbative estimates of $Z_q$
for different coupling constants, in the chiral limit and using two
renormalization conditions, for tree-level Symanzik gluons, $\csw=0,\,N_c=3$.
\end{minipage}
\end{center}}
\vspace{0.25cm}
The numbers is parenthesis denote the systematic errors coming from
our procedures for numerical integration over loop momenta.
Since we have used two alternative renormalization conditions,
it is worth comparing their lattice artifacts. For $\bar\mu^2=4$ and
the same parameters of Table 1, we isolated the contribution of the
${\cal O}(a^2)$ terms and tabulated them in Table 2. In the same Table
we present the averaged results, $\langle Z_q^{A\,{\cal O}(a^2)} \rangle,\,
\langle Z_q^{B\,{\cal O}(a^2)} \rangle$. For the averaging we used all
24 lattice momenta with $a^2p^2=4$: 16 permutations of
$(\pm1,\pm1,\pm1,\pm1)$ and 8 permutations of $(\pm2,0,0,0)$. One
observes that, although the condition in Eq.~(\ref{ZqA}) gives ${\cal O}(a^2)$
correction at tree level for $Z_q^A$, the combined ${\cal O}(a^2)$ 
effects in $Z_q^A$ are much smaller than the ones of $Z_q^B$, which is exactly 1
at tree level. It is also evident that averaging over momenta
with different directions helps to reduce the overall ${\cal O}(a^2)$
contributions.
\vspace{0.1cm}
\begin{center}
\begin{minipage}{12.5cm}
\begin{tabular}{lr@{}lr@{}lr@{}lr@{}lr@{}l}
\hline
\hline
\multicolumn{1}{c}{$\beta^{\phantom{A^{\phantom{A}}}}_{\phantom{A_{\phantom{A}}}}$}&
\multicolumn{2}{c}{$Z_q^{A\,{\cal O}(a^2)}$} &
\multicolumn{2}{c}{$\langle Z_q^{A\,{\cal O}(a^2)} \rangle$ } &
\multicolumn{2}{c}{$Z_q^{B\,{\cal O}(a^2)}$} &
\multicolumn{2}{c}{$\langle Z_q^{B\,{\cal O}(a^2)} \rangle$ } \\
\hline
\hline
3.80     &0&.35679267(9)    &0&.29105315(4)       &0&.45876742(9)    &0&.34204053(4)$^{\phantom{A^{\phantom{A}}}}$    \\
3.90     &0&.26014128(8)    &0&.23208860(4)       &0&.41544327(8)    &0&.30973960(4)                           \\
4.05     &0&.13958988(7)    &0&.15854326(3)       &0&.36140590(7)    &0&.26945127(3)$_{\phantom{A_{\phantom{A}}}}$    \\
\hline
\end{tabular}
\end{minipage}
\end{center}
{\small 
\centerline{{\bf Table 2:} The ${\cal O}(a^2)$ contributions up to 1-loop in the
perturbative results for the $Z_q$ at $\bar\mu^2=4$.}}

\section{${\cal O}(a^2)$ corrections to the renormalization of twist-2 fermion bilinears}
\label{twist2}

In this section we present the computation of the amputated Green's functions
for the following two twist-2 operators
\bea
{\cal O}^{\{\nu_1,\nu_2\}}_V &=& \frac{1}{2}\Big[\bar\Psi\,\gamma_{\nu_1}\,\Dlr\,\phantom{}_{\nu_2}\,\tau^\alpha\,\Psi + 
\bar\Psi\,\gamma_{\nu_2}\,\Dlr\,\phantom{}_{\nu_1}\,\tau^\alpha\,\Psi \Big] 
-\frac{1}{4} \delta_{\nu_1\nu_2} \sum_\rho \bar\Psi\,\gamma_\rho\,\Dlr\,\phantom{}_{\rho}\,\tau^\alpha\,\Psi  \\ 
{\cal O}^{\{\nu_1,\nu_2\}}_A &=& \frac{1}{2}\Big[\bar\Psi\,\gamma_5\gamma_{\nu_1}\,\Dlr\,\phantom{}_{\nu_2}\,\tau^\alpha\,\Psi + 
\bar\Psi\,\gamma_{\nu_2}\,\Dlr\,\phantom{}_{\nu_1}\,\tau^\alpha\,\Psi \Big] 
-\frac{1}{4} \delta_{\nu_1\nu_2} \sum_\rho \bar\Psi\,\gamma_\rho\,\Dlr\,\phantom{}_{\rho}\,\tau^\alpha\,\Psi 
\eea
which are symmetrized and traceless, to avoid mixing with lower
dimension operators. We have computed, to ${\cal O}(a^2)$, the matrix
elements of these operators for general external indices $\nu_1,\,\nu_2$,
mass, $g,\,N_c,\,a,\,p,\,\csw$ and $\lambda$. The final results are
available for the 10 sets of Symanzik coefficients we have used in
the calculation of $Z_q$.

The one-derivative operators fall into two different irreducible
representations of the hypercubic group, depending on the choice of
the external indices $\nu_1,\,\nu_2$. Hence, we distinguish between 
\be
Z^1_V \equiv Z^{\nu_1=\nu_2}_{V}\,, \qquad  Z^2_V \equiv Z^{\nu_1\neq\nu_2}_{V}\,, \qquad
Z^1_A \equiv Z^{\nu_1=\nu_2}_{A}\,,  \qquad Z^2_A \equiv Z^{\nu_1\neq\nu_2}_{A} 
\ee
The renormalization conditions from which $Z_V$ and $Z_A$ are obtained,
is defined in the RI'-MOM scheme, as
{\small{
\be
\label{ZO}
 Z_q^{-1} Z^{\nu_1\nu_2}_{\cal O} {\rm Tr} \Big[\Lambda^{\nu_1\nu_2}_{\cal O}(p)\, \cdot 
   \Lambda^{\nu_1\nu_2}_{{\cal O}\,cont} \Big]\Bigg{|}_{p^2 = \bar\mu^2}\hspace{-0.1cm} =  
{\rm Tr} \Big[\Lambda^{\nu_1\nu_2}_{{\cal O}\,cont} \cdot \Lambda^{\nu_1\nu_2}_{{\cal O}\,cont} \Big]\Bigg{|}_{p^2 = \bar\mu^2}, 
\quad \Lambda^{\nu_1\nu_2}_{{\cal O}\,cont} =
i\,\Gamma_{\{\nu_1}\,p_{\mu_2\}} - {\rm traces}
\ee
}}
\hskip -0.15cm
where $\Gamma_{\nu_1}=\gamma_{\nu_1},\,\gamma_5\,\gamma_{\nu_1}$.
The quantity $\Lambda^{\nu_1\nu_2}_{\cal O}(p)$ is our result up to
1-loop for the amputated Green's function for each operator. In the
renormalization condition there is a choice of using
$\Lambda^{\nu_1\nu_2}_{{\cal O}\,tree}$ instead of $\Lambda^{\nu_1\nu_2}_{{\cal O}\,cont}$. 
The final $Z$-factors of the two choices have different lattice artifacts.
Applying the renormalization condition and using the $Z_q$ that we
computed perturbatively with its ${\cal O}(a^2)$ corrections, we have
the results for $Z_V$ and $Z_A$. In Table 3 we present these results
in the chiral limit, for tree-level Symanzik gluons, $\beta=3.9$,
$\csw=0$ and Landau gauge. We chose six different values of the
momentum, thus six renormalization scales. For $Z_q$ we employed
Eq.~(\ref{ZqA}), and we used $\Lambda^{\nu_1\nu_2}_{{\cal O}\,cont}$
in Eq.~(\ref{ZO}). The extrapolation errors are smaller then the 
last digit shown in Table 3.
%
\begin{center}
\begin{minipage}{8cm}
\begin{tabular}{lr@{}lr@{}lr@{}lr@{}l}
\hline
\hline
\multicolumn{1}{c}{$a\,p^{\phantom{A^{\phantom{A}}}}_{\phantom{A_{\phantom{A}}}}$}&
\multicolumn{2}{c}{$Z_V^1$} &
\multicolumn{2}{c}{$Z_V^2$} &
\multicolumn{2}{c}{$Z_A^1$} &
\multicolumn{2}{c}{$Z_A^2$}  \\
\hline
\hline
(4,2,2,2)  &&1.04129   &&1.05738  &&1.08270 &&1.06309$^{\phantom{A^{\phantom{A}}}}$\\
(5,2,2,2)  &&1.04411   &&1.04331  &&1.08772 &&1.04669                       \\
(6,2,2,2)  &&1.04881   &&1.02365  &&1.09312 &&1.02496                       \\
(3,3,3,3)  &&1.00343   &&1.05427  &&1.03778 &&1.03607                       \\
(4,3,3,3)  &&0.99660   &&1.05237  &&1.04456 &&1.03289                       \\
(5,3,3,3)  &&0.99731   &&1.04736  &&1.05543 &&1.02660$_{\phantom{A_{\phantom{A}}}}$
\\
\hline
\end{tabular}
\end{minipage}
\end{center}
{\small 
\begin{center}
\begin{minipage}{14cm}
{\bf Table 3:} Perturbative results for $Z_A$ and $Z_V$ in the chiral
limit, for various renormalization scales. The action parameters are:
tree-level Symanzik gluons, Landau gauge, $\beta=3.9,\,\csw=0,\,N_c=3$.
\end{minipage}
\end{center}}
For the case of the vector operator with  $\nu_1\neq\nu_2$ we give the
result for ${\rm Tr} \Big[\Lambda^{\nu_1\nu_2}_V(p)\, \cdot \Lambda^{\nu_1\nu_2}_{{\cal O}\,cont}\Big]$
which contributes to $Z_V^2$, for the Landau gauge, tree-level
Symanzik gluons, $\csw=0,\,\mu=0$
\vspace{0.5cm}
{\small{
\bea
\hspace{-0.5cm}
{\rm Tr} \Big[\Lambda^{\nu_1\nu_2}_V(p)\, \cdot \Lambda^{\nu_1\nu_2}_{{\cal O}\,cont}\Big]&=&
-p_{\nu_1}^2 -p_{\nu_2}^2 + \Orange{a^2} \frac{p_{\nu_1}^4+p_{\nu_2}^4}{6} 
+ \frac{C_F\,g^2}{16\,\pi^2} \Red{\Bigg[}
  \,15.045752(1)\,p_{\nu_1}^2 + \frac{4\,p_{\nu_1}^2\,p_{\nu_2}^2}{3\,p^2} \nonumber \\
&-& \frac{8\,p_{\nu_1}^2\,\ln(p^2)}{3} 
+ \Orange{a^2}\,\Green{\Bigg(} -0.13212(3)\,p^2\,p_{\nu_1}^2 -
     4.6352(1)\,p_{\nu_1}^4 - 4.0096(1)\,p_{\nu_1}^2\,p_{\nu_2}^2 \nonumber \\ 
&+& \frac{353}{720}\,\frac{p4\,p_{\nu_1}^2}{p^2} 
+  \frac{29}{90}\,\frac{p4\,p_{\nu_1}^2\,p_{\nu_2}^2}{(p^2)^2} +
     \frac{179}{90}\,\frac{(p_{\nu_1}^4\,p_{\nu_2}^2 + p_{\nu_1}^2\,p_{\nu_2}^4)}{p^2}\nonumber \\ 
&+& \ln(a^2\,p^2) \Big(-\frac{103}{360}\,p^2\,p_{\nu_1}^2  + \frac{331}{360}\,p_{\nu_1}^4  + 
     \frac{1013}{180}\,p_{\nu_1}^2\,p_{\nu_2}^2 \Big)\Green{\Bigg)}\nonumber \\ 
&+& {\rm symmetric\,\, terms:\,\,}{\nu_1} \leftrightarrow {\nu_2} \Red{\Bigg]}
\eea}}
It is now interesting to use non-perturbative estimates for $Z_A$ and
$Z_V$, combined with our ${\cal O}(a^2)$ terms, in order to see if the
subtraction procedure works well for these operators. Our
collaborators in ETMC provided us with data for the physical
$Z_A^2$~\cite{ACK} corresponding to the six momenta of Table 3. Note
that the momenta can be grouped into two categories, according to
their spatial values: (2,2,2) or (3,3,3). Non-perturbatively it is observed that these
groups have different lattice artifacts and a discontinuity appears in
the $Z$-factors with one derivative, as shown in Fig. 2. We also plot
the subtracted non-perturbative data, which exhibit a smooth behavior of $Z_A^2$.
The errors are too small to be visible.
\vspace{-1cm}
\begin{center}
\psfig{figure=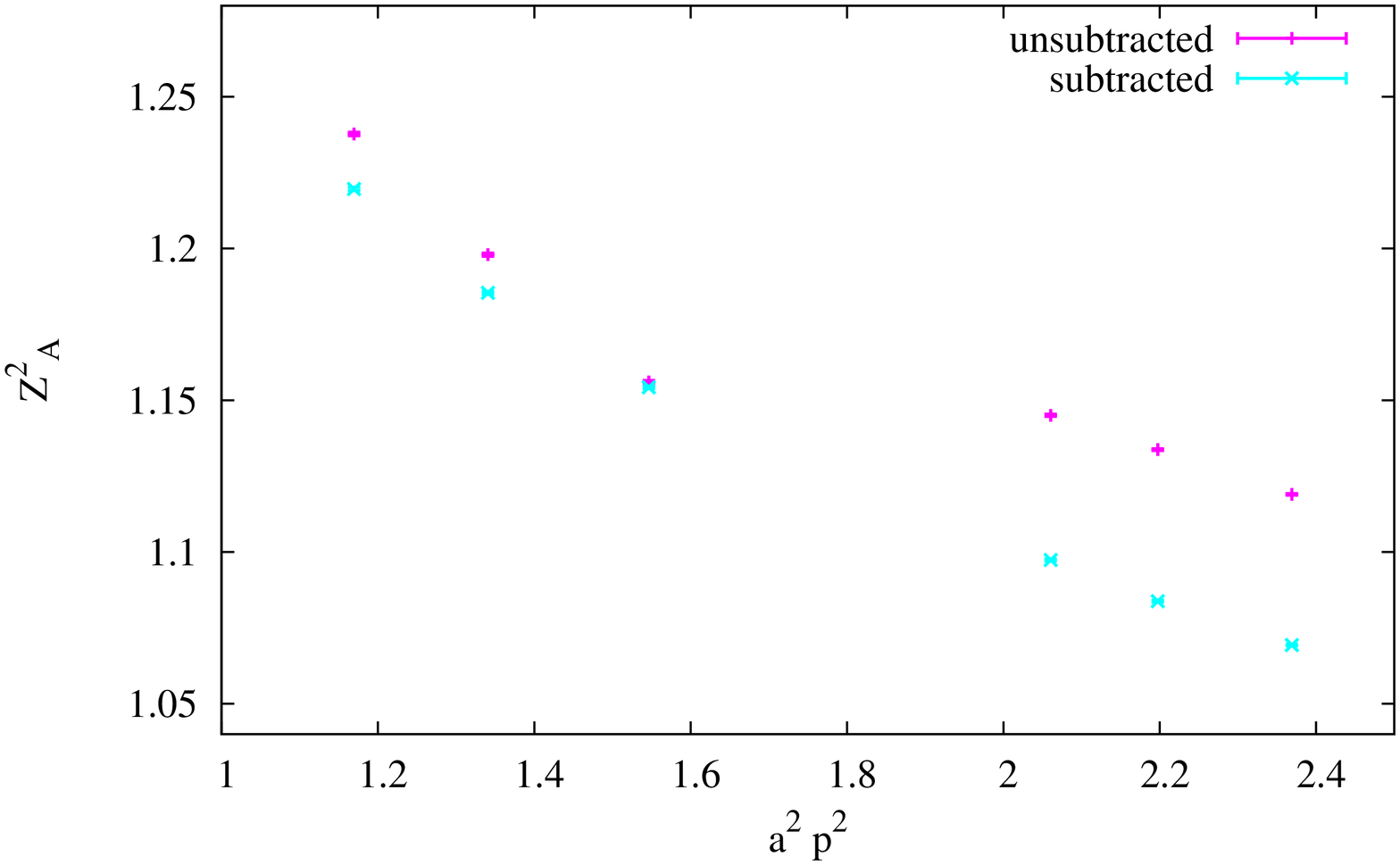,height=10truecm,angle=-90}
\end{center}
{\small
\centerline{{\bf Figure 2:} Physical non-perturbative $Z_A^2$ with subtractions of
${\cal O}(a^2)$ terms.}}
\vspace{0.5cm}

Another plot demonstrating the achieved improvement of
non-perturbative estimates for the renormalization constants using our
$a^2$ correction terms, is shown in Ref.~\cite{ACK2}. In particular,
for the local axial renormalization constant it is shown that, by
converting the subtracted numbers into $\overline{MS}$ scheme and setting the
renormalization scale to 2GeV, one obtains a good plateau when
plotting $Z_A$ for each of the momentum 4-vectors.


\end{document}